\documentclass[12pt]{article}
\pdfoutput=1
\usepackage{hyperref}
\usepackage{graphicx}
\usepackage{graphics}
\usepackage{dsfont}
\usepackage{epsfig}
\usepackage{amsmath,amssymb,amsthm,amscd}

\newcommand{\Res}{\textrm{Res}}
\def\ket{\rangle}
\def\bra{\langle}

\newcommand{\be}{\begin{equation}}
\newcommand{\ee}{\end{equation}}
\newcommand{\ba}{\begin{aligned}}
\newcommand{\ea}{\end{aligned}}

\numberwithin{equation}{section}

\begin{document}
\begin{titlepage}
{}~ \hfill\vbox{ \hbox{} }\break

\rightline{USTC-ICTS/PCFT-20-30}

\vskip 3 cm

\centerline{\Large 
\bf  
Quantum Periods  and TBA-like Equations } 
\vskip 0.2 cm
\centerline{\Large 
\bf 
for a Class of Calabi-Yau Geometries  }

\vskip 0.5 cm

\renewcommand{\thefootnote}{\fnsymbol{footnote}}
\vskip 30pt \centerline{ {\large \rm 
Bao-ning Du\footnote{baoningd@mail.ustc.edu.cn}, 
Min-xin Huang\footnote{minxin@ustc.edu.cn}  
} } \vskip .5cm  \vskip 20pt 

\begin{center}
{Interdisciplinary Center for Theoretical Study,  \\ \vskip 0.1cm  University of Science and Technology of China,  Hefei, Anhui 230026, China} 
 \\ \vskip 0.3 cm
{Peng Huanwu Center for Fundamental Theory,  \\ \vskip 0.1cm  Hefei, Anhui 230026, China} 
\end{center}

\setcounter{footnote}{0}
\renewcommand{\thefootnote}{\arabic{footnote}}
\vskip 60pt
\begin{abstract}
We continue the study of a novel relation between quantum periods and TBA(Thermodynamic Bethe Ansatz)-like difference equations, generalize previous works to a large class of Calabi-Yau geometries described by three-term quantum operators. We give two methods to derive the TBA-like equations. One method uses only elementary functions while the other method uses  Faddeev's quantum dilogarithm function. The two approaches provide different realizations of TBA-like equations which are nevertheless related to the same quantum period.

\end{abstract}

\end{titlepage}
\vfill \eject


\newpage

\baselineskip=16pt

\tableofcontents

\section{Introduction}

Period integrals are ubiquitous in mathematics and physics. For example, they are essential ingredients in classic mirror symmetry \cite{Candelas:1990}, in the solutions of Seiberg-Witten gauge theories \cite{Seiberg:1994rs}, and are also frequently related to appealing mathematical objects like modular forms \cite{zagier2008elliptic}. They even appear in some phenomenologically relevant topics, e.g. in Feynman diagram calculations \cite{Bonisch:2020qmm}. On the other hand, Bethe Ansatz has served as a cornerstone in the developments of exactly solvable many-body quantum systems for almost a century. We will consider a type of difference equations similar to those that appear in the Thermodynamic Bethe Ansatz (TBA).    

Quantum periods have been a useful tool in recent studies of topological string theory and related topics. We will consider local Calabi-Yau geometries which can be described by a complex one-dimensional mirror curve with complex coordinates $(x,p)$. Promoting the coordinates to canonical quantum position and momentum operators, we can then compute the quantum periods as integrals of a quantum corrected differential one-form over cycles, see e.g. \cite{Aganagic}. In this paper, we will consider the case of simple cycles where the calculations of period integral reduce to a residue. These cycles are usually known as A-cycles. In Calabi-Yau geometries, the classical A-period is the mirror map between from complex structure moduli of a Calabi-Yau space to the Kahler moduli of the mirror Calabi-Yau. So the quantum A-period is also known as the quantum mirror map. While the direct computations of the quantum periods over the other complicated cycles are more difficult, the quantum corrections are determined by the same differential operators as the quantum A-period, see e.g. \cite{Mironov:2009, Huang:2012}. These developments in quantum periods lead to exciting results such as the calculations of topological string free energy in the NS limit \cite{Nekrasov:2009}, quantum spectra \cite{Huang:2014, Huang:2020neq}, exact quantizations including non-perturbative effects \cite{Grassi, Wang:2015wdy}. The quantum mirror maps for a class of del Pezzo Calabi-Yau geometries with exceptional symmetry are recently studied in \cite{Moriyama:2020lyk}.

We will study the novel relation between quantum periods and TBA-like equations proposed in \cite{Hatsuda:2013oxa, Kallen:2013qla}. In particular, we focus on the parts concerning quantum A-periods which can be determined by a residue calculation at finite Planck constant $\hbar$. The original proposal is for a particular local $\mathbb{P}^1\times \mathbb{P}^1$ Calabi-Yau geometry related to the ABJM (Aharony-Bergman-Jafferis-Maldacena) theory \cite{ABJM} and has been generalized recently by one of the authors \cite{Huang:2020yli} to include the mass parameter in the Calabi-Yau geometry. The case of local $\mathbb{P}^2$ model was also studied in \cite{Okuyama:2015pzt}. Some related papers leading to the relation include \cite{Marino:2011eh, Calvo:2012du, Hatsuda:2012hm, Putrov:2012zi}, starting from e.g. early papers \cite{Zamolodchikov:1994uw, Tracy:1995ax}. In this paper, we will further explore the relation and generalize to a large class of local Calabi-Yau geometries. 

We should note that in our study we only use the TBA-like difference equations and will not need integral equations which may be more familiar for obtaining the analytic properties of the underlying quantum spectral problem. In this paper we will not study the spectral theory but only solve the difference equations perturbatively around a small complex structure deformation parameter for the purpose of comparing with quantum A-periods. So our difference equations are similar to, but may not be equivalent to those of an actual TBA system. In this sense, they are usually called TBA-like. 

This paper is organized as the followings. In Section \ref{sec2} we review some general formalism. We will then use the general formalism to derive the proposal in the previous paper \cite{Huang:2020yli} for TBA-like equations for the $\mathbb{P}^1\times \mathbb{P}^1$ geometry, including the mass parameter. In Section \ref{sec3}, we consider the class of local Calabi-Yau geometries described by three-term operators. We compute the quantum periods for a one-parameter deformation in the moduli space. It turns out that the quantum mirror curves can be transformed by some elementary manipulations to a form so that the general formalism can apply. We write down the corresponding TBA-like equations in this way and check the relations with quantum periods. In Section \ref{sec4}, we take another approach that expresses the operators in terms of Faddeev's quantum dilogarithm. We then follow the literature to derive the corresponding TBA-like equations. It turns out the final equations also involves only elementary functions due to some useful formulas of the quantum dilogarithm. In the final Section \ref{sec5}, we discuss some potential applications, intriguing features of our results and future directions.

\section{Quantum periods and TBA-like equations} \label{sec2}

In this section, we shall briefly review the general derivation of the relation between quantum period and TBA-like equation as well as the specific application to the local $\mathbb{P}^1\times \mathbb{P}^1$ model, which will provide a useful ground for further generalizations. 

The classical mirror curves of local Calabi-Yau geometries can be quantized by promoting the two complex coordinates of the curves to quantum position and momentum operators with the canonical commutation relation $[\hat{x}, \hat{p}]=i\hbar$. Acting the quantized mirror curve on a wave function usually gives a difference equation, due to the appearance of the exponential $e^{\hat{p}}$. We can then compute the quantum A-period $\Pi_A$ by a simple residue calculation, see e.g. \cite{Aganagic}. The quantum B-period involves cycle integrals over to infinity and is generally much more complicated to do than the quantum A-period. The quantum periods depend on the complex structure moduli of the mirror curves as well as the Planck constant $\hbar$. 

The quantization of the mirror curve defines a quantum spectral problem, as we parametrize the dynamical complex structure moduli in terms of Hamiltonians and separate them out.  In this paper we consider Calabi-Yau geometries or a subfamily of its moduli space with only one dynamical modulus. In this case, after some changes of variables, we can write the spectral equation as 
\begin{eqnarray} \label{spectral} 
\rho(\hat{x}, \hat{p}) |\psi\ket = e^{-E}  |\psi\ket, 
\end{eqnarray} 
where the energy $E$ is related to the only dynamical complex structure modulus. The spectral operator is a Hermitian trace class operator with discrete eigenvalues. For the purpose of deriving the corresponding TBA-like equation,  it needs to have the following form 
\begin{eqnarray} \label{spectraloperator}
\rho(\hat{x}, \hat{p}) =  u(\hat{x})^{-\frac{1}{2}}   [2\cosh(\hat{p})]^{-1}  u^*(\hat{x})^{-\frac{1}{2}}, 
\end{eqnarray} 
where the function $u(x)$ may depends on non-dynamical mass parameters of the Calabi-Yau geometries. The matrix element in terms of position eigenstates can be computed by a simple Fourier transform of the $\cosh(p)^{-1}$ function 
\be
\bra x_1 | \rho | x_2\ket = \frac{ u(x_1)^{-\frac{1}{2}}  u^*(x_2)^{-\frac{1}{2}} }{ 4\hbar \cosh(\frac{\pi(x_1-x_2)}{2\hbar} )}
\ee

In the 1990's, it was conjectured in \cite{Zamolodchikov:1994uw} and proved in \cite{Tracy:1995ax} that the spectral equation (\ref{spectral}) is related to some TBA-like equations which first appeared in the context of two-dimensional $\mathcal{N} = 2$ supersymmetric theories \cite{Fendley:1992jy, Cecotti:1992qh}. The integral form of the TBA-like equations can be transformed to a difference equation by a Fourier transform, e.g. for the case of ABJM theory in \cite{Calvo:2012du}. We provide some details of the derivations here. We use the variable $\theta \equiv \frac{\pi x}{\hbar}$ which is common in the literature and absorbs the Planck constant. We also define a function $U(x) \equiv  \log(|u(x)|)$, since we can always perform a unitary transformation to absorb the phase in in the function $u(x)$, which does not change the spectrum. So the spectral equation (\ref{spectral}) is equivalent to the study of the following integral kernel 
\be
K(\theta, \theta^{\prime}) =\frac{1}{2\pi} \frac{\exp[-\frac{U(\theta)+U(\theta^{\prime})}{2} ]} {2\cosh(\frac{\theta-\theta^{\prime}}{2})}.  
\ee
One introduces  $R_0(\theta)=e^{-U(\theta)}$ and the iterated integral 
\be
R_l(\theta) = e^{-U(\theta)} \int_{-\infty}^{\infty} \frac{\exp[ -\sum_{i=1}^lU(\theta_i) ]}{ \cosh\frac{\theta-\theta_1}{2} 
\cosh\frac{\theta_1-\theta_2}{2}  \cdots \cosh\frac{\theta_l-\theta}{2}  } d\theta_1 \cdots d\theta_l, ~~~ l\geq 1. 
\ee
Furthermore we define a generating series and the decomposition into the even and odd parts 
\be
\ba
& R(\theta| z) =\sum_{l=0}^{\infty} (-\frac{z}{4\pi} )^l R_l(\theta) ,  \\
& R_+(\theta| z) = \frac{1}{2} [ R(\theta| z) + R(\theta| -z)], ~~~  R_-(\theta| z) = \frac{1}{2} [ R(\theta| z) - R(\theta| -z)]. 
\ea
\ee
Then the theorems of \cite{Zamolodchikov:1994uw, Tracy:1995ax} state that the even and odd  generating functions can be computed by a TBA system with the integral equations 
\be \label{integralequation}
\ba
U(\theta) &= \epsilon(\theta) +\int_{-\infty}^{\infty} \frac{d\theta^{\prime}}{2\pi} \frac{\log(1+\eta^2(\theta^{\prime}))}{\cosh(\theta-\theta^{\prime})} , \\
\eta(\theta) &=-z \int_{-\infty}^{\infty}  \frac{d\theta^{\prime}}{2\pi} \frac{e^{-\epsilon(\theta^{\prime})} }{\cosh(\theta-\theta^{\prime})}.
\ea
\ee
The TBA integral equations determine the two functions $\eta(\theta), \epsilon(\theta)$ and we have 
\be
\ba
 R_+(\theta| z) = e^{-\epsilon(\theta)}, ~~~~  R_-(\theta| z) = R_+(\theta| z) \int_{-\infty}^{\infty} 
  \frac{d\theta^{\prime}}{\pi} \frac{\arctan(\eta(\theta^{\prime})) }{\cosh^2(\theta-\theta^{\prime})}.
\ea
\ee
To derive the difference equations, we perform Fourier transforms on both sides of the two integral equations (\ref{integralequation}). We use the integral formula $\int_{-\infty}^{\infty}\frac{e^{i \theta \xi} d\theta }{\cosh \theta} = \frac{\pi}{\cosh(\frac{\pi \xi}{2})} $, where $\xi$ is the Fourier conjugate variable of $\theta$. We then multiply both sides by $\cosh(\frac{\pi \xi}{2})$ and perform the inverse Fourier transforms back to the $\theta$ variable,  arriving at the following difference equations  
\be
\ba
U(\theta+\frac{\pi i}{2})+ U(\theta-\frac{\pi i}{2}) &= \epsilon(\theta+\frac{\pi i}{2}) +\epsilon(\theta - \frac{\pi i}{2})  
+\log(1+\eta^2(\theta)), \\
\eta(\theta+\frac{\pi i}{2}) + \eta(\theta - \frac{\pi i}{2}) &=-ze^{-\epsilon(\theta)}  . 
\ea
\ee
We can eliminate the function $\epsilon(\theta)$ and obtain a difference equation for $\eta(\theta)$. In order to compare with quantum periods, we change back the variable $\theta=\frac{\pi x}{\hbar}$ and also need to make some transformations $z\rightarrow z^{-\frac{1}{2}}, \eta\rightarrow i\eta$. We will use the TBA-like difference equation for the function $\eta(X)$ after these manipulations
\begin{eqnarray} \label{TBAgeneral} 
1+z[\eta(qX) +\eta(X)] [ \eta(q^{-1} X) +\eta(X)] |u(q^{\frac{1}{2}} X)u(q^{-\frac{1}{2}} X)| =\eta(X)^2 , 
\end{eqnarray} 
where we use the exponentiated parameters $X=e^x, q=e^{i\hbar}$.  We will study the relation proposed more recently in \cite{Hatsuda:2013oxa} that the derivative of quantum A-period $z\partial_z \Pi_A$ is related to the constant term of $X$ in the Laurent expansion of $\eta(X)$, or equivalently a residue of $x=\log(X)$. The $z$ parameter in (\ref{TBAgeneral}) will be identified with the complex structure moduli parameter. We note that the function $u(x)$ is defined in (\ref{spectraloperator}) up to a constant factor which can come from a shift of the energy $E$ in (\ref{spectral}).  Since our derivation here is only motivational but not completely rigorous, in practice, the  determination of such constant factor as well as the precise relation of the quantum A-period with the residue from the TBA-like equation will be obtained by comparing the results for specific Calabi-Yau geometries. In this paper we focus on toric CY manifolds where the explicit expression of the inverse kernel is known and is of the form (\ref{spectraloperator}), otherwise it is not clear to us whether the analogous TBA-like equation like (\ref{TBAgeneral}) can be obtained.

Let us review the case of local $\mathbb{P}^1\times \mathbb{P}^1$ model, which has the mirror curve 
\be
e^x+e^p+z_1 e^{-x} + z_2 e^{-p} =1. 
\ee
Promote the $x,p$ coordinates to quantum operators and consider a particular case $z_1=q^{-\frac{1}{2} } z, z_2=q^{\frac{1}{2} } z$ which corresponds to the ABJM theory \footnote{In the previous paper \cite{Huang:2020yli},   we use a different parametrization  $z_1=q^{\frac{1}{2} } z, z_2=q^{-\frac{1}{2} } z$, which gives the same result for quantum A-period. However, the present parametrization is suitable for writing the spectral equation in the form of (\ref{spectral}, \ref{spectraloperator}) in terms of elementary functions. }. We make a change of variables 
\be
\hat{x} \rightarrow \frac{\hat{x}}{2} -\hat{p} - \frac{i\hbar}{4} -E, ~~\hat{p} \rightarrow \frac{\hat{x}}{2} +\hat{p} +\frac{i\hbar}{4} -E, ~~z\rightarrow e^{-2E}, 
\ee  
which preserve the same canonical commutation relation. Acting the mirror curve on a quantum state $|\phi\ket$ we get
\be
4\cosh(\frac{\hat{x}}{2}) \cosh(\hat{p}) |\phi\ket = e^{E} |\phi\ket. 
\ee 
Finally we make a change of quantum state $|\phi\ket  = (e^{\frac{\hat{x}}{2}} + e^{-\frac{\hat{x}}{2} })^{\frac{1}{2}} |\psi\ket$ so that we have a spectral equation in the form of (\ref{spectral}, \ref{spectraloperator}), with the function $u(x)= e^{\frac{x}{2}} + e^{-\frac{x}{2} } $. This provides the corresponding TBA-like equation for the ABJM theory by plugging the function into (\ref{TBAgeneral}).

The case of general $\mathbb{P}^1\times \mathbb{P}^1$ model where the complex structure parameters include the mass parameter  $z_1=e^{\tilde{m}} z, z_1=e^{- \tilde{m} } z$  is considered in \cite{Huang:2020yli}, and the corresponding TBA-like equation is proposed. Here we use a tilde symbol for the mass parameter $\tilde{m}$ to distinguish from the integer in the next sections. However, unlike the particular case $z_1=q^{-\frac{1}{2} } z, z_2=q^{\frac{1}{2} } z$ for ABJM theory, in the general case it seems difficult to write the spectral equation in the form of (\ref{spectral}, \ref{spectraloperator}) by simple manipulations of elementary functions. Instead, we will need to use Faddeev's  quantum dilogarithm function, defined in \cite{Faddeev:1993rs}. We list some useful properties in Appendix \ref{Qdilogorithm}. The quantum operator with general mass parameter can be written in the form of of (\ref{spectral}, \ref{spectraloperator}), as described in \cite{Kashaev:2015, Marino:2015ixa, Kashaev:2015wia}. We quote the result in \cite{Kashaev:2015wia} in terms of our current notation 
\be
\rho  = (e^{\hat{x}}+e^{\hat{p}}+e^{\tilde{m}} e^{-\hat{x}} + e^{-\tilde{m}} e^{-\hat{p}} )^{-1} =  f(\hat{x^{\prime}})  [2\cosh(\hat{p^\prime})]^{-1} f^*(\hat{x^{\prime}}) ,  
\ee
where the redefined variables are 
\be
\hat{x^{\prime} }= \hat{x}- \hat{p} - \tilde{m}, ~~~ \hat{p^{\prime} }= \frac{1}{2}(\hat{x}+ \hat{p} - \tilde{m}), 
\ee
which satisfy the same canonical commutation relation. The function $f(x)$ is defined in terms of the quantum dilogarithm
\be
f(x) = e^{\frac{x}{4}} \Phi_b (\frac{x-\tilde{m}}{2\pi b} +i\frac{b}{4})  \Phi_b (\frac{x+\tilde{m}}{2\pi b} - i\frac{b}{4})^{-1} , 
\ee 
with the parameter definition $b=\sqrt{\frac{\hbar}{\pi}}$. Using the well known functional relations of quantum dilogarithm, it is straightforward to calculate 
\be
|f(x+\frac{i\hbar}{2}) f(x- \frac{i\hbar}{2}) |^{-2} = e^x+e^{-x} +e^{\tilde{m}} +e^{-\tilde{m}} . 
\ee
So we can now derive the TBA-like equation for $\mathbb{P}^1\times \mathbb{P}^1$ model with general mass parameter 
\be \label{massP1P1}
1+z[\eta(qX) +\eta(X)] [ \eta(q^{-1} X) +\eta(X)] (X+1/X +e^{\tilde{m}} +e^{-\tilde{m}} )=\eta(X)^2 , 
\ee
which confirms the proposal in the previous paper \cite{Huang:2020yli}.

\section{The  $\mathcal{O}_{m,1}$ operators: first method}   \label{sec3}

We consider a class of three-term quantum operators of the form 
\be \label{OMN}
\mathcal{O}_{m,n} = e^{\hat{x}} +e^{\hat{p}} + e^{-m \hat{x} -n \hat{p}} , 
\ee 
which were also studied in \cite{Kashaev:2015}. Here $m,n$ are natural numbers and in this section we will focus on the case of $n=1$ where the corresponding TBA-like equation can be derived by the elementary method in this section. The case of $m=n=1$ corresponds to the well studied local  $\mathbb{P}^2$ Calabi-Yau geometry, while the case of $(m,n)=(2,1)$ corresponds to a subfamily of  local Hirzebruch $\mathbb{F}_2$ Calabi-Yau geometry. In general, the operator corresponds to a $\mathbb{C}^3/\mathbb{Z}_{m+2}$ resolved orbifold Calabi-Yau space. For $m>2$, the operator has multiple complex deformations which correspond to different dynamical Hamiltonians, we will consider a one-parameter subfamily of such deformations. 

The quantum mirror curve of the $\mathcal{O}_{m,1}$ operator can be parametrized as 
\be \label{mncurve} 
e^{\hat{x}}+e^{\hat{p}}+ze^{-m \hat{x}-\hat{p}} =1 .
\ee
The calculation of the quantum A-period is now well known, see e.g. \cite{Aganagic}. We act the quantum curve on a wave function $\psi(x)$ and denote $V(x)= \frac{\psi(x)}{\psi(x-i\hbar)}$, then we have a difference equation 
\be \label{quadratic3.3}
X-1+\frac{1}{V(X)}  +\frac{zV(qX)}{q^{\frac{m}{2}}X^m} =0,
\ee
where again the notation is $X=e^x, q=e^{i\hbar}$. This equation is quadratic in the classical limit $q\rightarrow 1$. Without loss of generality, we choose one of the two solutions which is regular in the $z\rightarrow 0$ limit. The difference equation for $V(X)$ can then be solved perturbatively as a power series in $z$, and for example up to order $z$ we have the expression  
\be
V(X)= \frac{1}{1-X} +\frac{q^{-\frac{m}{2}}X^{-m} z}{(1-X)^2(1-qX)} + \mathcal{O}(z^2).  
\ee
We consider the residue around $X=0$  
\be
\Pi_m = \Res_{X=0} \frac{\log[V(X)] }{X} =  \Res_{X=0} \frac{\log[ (1-X) V(X)] }{X}. 
\ee
We compute the residue and list some explicit expressions for small $m$'s 
\be \label{Pim}
\ba
\Pi_1 &=  \frac{ (1 + q) z}{\sqrt{q}} + \frac{ [2(1+q^4) + 7 (q+q^3) + 12 q^2 ] z^2}{ 2 q^2} + [3(1+q^9) + 9( q+q^8) 
 \\  & + 36 (q^2+q^7)  + 88 (q^3 +q^6) + 144 (q^4+q^5) 
] \frac{z^3}{3q^{9/2}} +\mathcal{O}(z^4),  \\
\Pi_2 &= \frac{ (1 + q+q^2) z}{q} + [\frac{27}{2} +10(q+q^{-1}) +\frac{13}{2} (q^2+q^{-2}) +2(q^3+q^{-3}) 
+q^4+q^{-4}  ]z^2  \\ & +\mathcal{O}(z^3),   \\
\Pi_3 &= \frac{ (1 + q+q^2+q^3) z}{q^{\frac{3}{2}}} + [24 +\frac{39}{2} (q+q^{-1}) +15 (q^2+q^{-2}) +\frac{19}{2}(q^3+q^{-3}) \\ &  +4(q^4+q^{-4} ) +2 (q^5+q^{-5}) +q^6+q^{-6}  ]z^2  +\mathcal{O}(z^3),    \\
\Pi_4 &= \frac{ (1 + q+q^2+q^3+q^4) z}{q^2} + [\frac{75}{2} +10(q+q^{-1}) +
32( q+q^{-1}) +\frac{53}{2} (q^2 +q^{-2}) \\ & +20 (q^3+q^{-3}) 
+\frac{27}{2} (q^4+q^{-4}) +6 (q^5+q^{-5}) +4 (q^6 +q^{-6}) +2 (q^7+q^{-7})  \\ &  +q^8 +q^{-8}   ]z^2 +\mathcal{O}(z^3), 
\ea
\ee

For the cases of $m=1,2$, the small $z$ limit represents the large volume points in the local $\mathbb{P}^2$ geometry and the local $\mathbb{F}_2$ geometry with a special mass parameter. We need to add an appropriate $\log(z)$ term which is well known to present in the classical A-period but not captured by the residue calculations here. So we can write the quantum A-period as 
\be
\Pi_{\mathbb{P}^2, A} =   \log(z) +3\Pi_1, ~~~ \Pi_{\mathbb{F}_2, A} =   \log(z) + 4\Pi_2 . 
\ee 
Here the proper combinations can be found by comparing with the classical period, are also available  in e.g. \cite{Wang:2015wdy}, with the mass parameter in $\mathbb{F}_2$ geometry set to zero. 

However, for the cases of $m>2$, the small $z$ limit may not be the large volume point of the Calabi-Yau geometry. For example, consider the case of $m=3$, a parametrization of the resolved $\mathbb{C}^3/\mathbb{Z}_{5}$ mirror curve can be found in e.g. \cite{Franco:2015rnr} as
\be
e^{\hat{x}} +e^{\hat{p}}+e^{-3\hat{x}-\hat{p}} 
+x_0 e^{-\hat{x}} +x_3 = 0,
\ee
where $x_0,x_3$ are related to the Batyrev complex structure coordinates by 
\be
z_1 = \frac{x_3}{x_0^3 }, ~~~ z_2 = \frac{x_0}{x_3^2}. 
\ee
The large volume point is $(z_1,z_2)\sim (0,0)$. On the other hand, we can relate to the curve (\ref{mncurve}) by setting $x_0=0$ and shift the operators $\hat{x}, \hat{p}$. The parameter in  the curve (\ref{mncurve}) is then related  $z = x_3^{-5}$. We see that the small $z\sim 0$ limit here is not the large volume point $(z_1,z_2)\sim (0,0)$. In other points of the moduli space such as conifold and orbifold points, the classical mirror map is regular with no logarithmic behavior, so in those cases we will not need to add a $\log(z)$ term to obtain the quantum A-period. In the followings, we will simply use the residue $\Pi_m$, which is actually more convenient for the purpose of comparing with the calculations from the corresponding TBA-like equations.

To derive the corresponding TBA-like equation, we change variables 
\be
\hat{x} \rightarrow \hat{x} , ~~ \hat{p}\rightarrow -m \frac{\hat{x}}{2} +\hat{p} - \frac{im \hbar}{4} -E, ~~ z\rightarrow e^{-2E}. 
\ee
Acting the mirror curve on a quantum state $|\phi\ket$, we find 
\be
2\cosh(\hat{p}) |\phi\ket =e^{ \frac{m \hat{x}}{2}}  (1- e^{\hat{x}} )e^E |\phi\ket . 
\ee
After a simple redefinition of the quantum state, we can now write the spectral equation in the form of  (\ref{spectral}, \ref{spectraloperator}) with the function $u(x) = e^{-\frac{mx}{2}} (1-e^x)^{-1}$. 

We should note that unlike the examples in  \cite{Kashaev:2015},  in our case the integral $\int_{-\infty}^{\infty} |u(x)|^{-1} dx$ is actually divergent, which implies the corresponding integral kernel (\ref{spectraloperator}) may not be a trace class operator. However this is not really an issue since we are not studying the spectral theory, but just use it as a formal trick to derive the TBA-like equations. Our end result should justify ignoring such subtleties in the process. 

So using (\ref{TBAgeneral}), we arrive at a TBA-like equation for the $\mathcal{O}_{m,1}$ operator
\be \label{firstTBA}
1+z[\eta_m(qX) +\eta_m(X)] [ \eta_m(q^{-1} X) +\eta_m(X)]/[X^{m+1} (X+X^{-1} -q^{\frac{1}{2}} - q^{-\frac{1}{2}})]  =\eta_m(X)^2 .
\ee
Here since $X=e^x>0$ and we will take residue around $X\sim 0$, we can assume the function always has definite sign $X+X^{-1} -q^{\frac{1}{2}} - q^{-\frac{1}{2}}>0$ in the absolute value in (\ref{TBAgeneral}). We can solve this difference equation for $\eta_m(X)$ as a perturbative series of $z$. With the plus sign leading term, we find  e.g, up to order $z$ term
\be
\eta_m(X, q, z) = 
1 + \frac{2  z}{X^{m+1} (X+X^{-1} -q^{\frac{1}{2}} - q^{-\frac{1}{2}}) } + \mathcal{O}(z^2). 
\ee
We take the residue and check up to the first few orders that it is indeed simply related to the residue (\ref{Pim}) in quantum periods for some small numbers $m$ by the following  
\be \label{compare1}
\Res_{X=0} \frac{1}{X} \eta_m(X, q, z) = 1+2  \theta_z \Pi_m  ,
\ee
where $\theta_z \equiv z\partial_z$. Unlike the case of $\mathbb{P}^1\times \mathbb{P}^1$ model, here we have some non-trivial constants in the relation. Since we do not give a rigorous proof of the relation, these constants  are determined here on a rather ad hoc basis by checking the results. 

Although in this paper we treat the $z$ expansion as a formal power series, we can comment on the convergence of the series. Since the residues of $\eta_m$'s are related to the quantum A-periods, their convergence behaviors are the same so we only need to look at the series in (\ref{Pim}). As familiar in many examples, the quantum corrections to the periods can be obtained by some differential operators acting on the classical periods, so the convergence behaviors of quantum periods should be the same as the classical periods which are solutions of the Picard-Fuchs differential equations. According to the general theory, the expansion of the classical period is convergent until we touch the nearest singular point in the moduli space. For example, for the case of $m=1$, the nearest singular point is a conifold point of the local $\mathbb{P}^2$ Calabi-Yau geometry at $z=\frac{1}{27}$, so in this case the series $\Pi_1$ in (\ref{Pim}) should have a convergence radius $\frac{1}{27}$. For the other cases $m>1$, one needs to study carefully the moduli space, but we expect there is always a positive finite radius of convergence.

The case $m=2$ is somewhat interesting, as it is known that the $\mathbb{F}_2$ model is related to the $\mathbb{F}_0\equiv  \mathbb{P}^1\times \mathbb{P}^1$ model by a reparametrization. The relation of parameters can be found in e.g. \cite{Wang:2015wdy}. Basically, our case corresponds to setting the mass parameter $\tilde{m}=\frac{\pi i}{2}$ in the TBA-like equation (\ref{massP1P1}) for the $\mathbb{P}^1\times \mathbb{P}^1$ model. There is also a rescaling of parameter $z\rightarrow z^{\frac{1}{2}}$ due to different scalings with respect to the energy in the parametrization of the curves. So we have another TBA-like equation for the $m=2$ case as 
\be \label{F2equation}
1+z^{\frac{1}{2}}[\tilde{\eta}(qX) + \tilde{\eta} (X)] [ \tilde{\eta} (q^{-1} X) + \tilde{\eta}(X)] (X+1/X )=\tilde{\eta} (X)^2 . 
\ee
The perturbative solution for $\tilde{\eta}$ has half integer powers of $z$. However after taking residue, only the integer powers survive. We check that it is indeed simply related to $\eta_2$ by the following intriguing equation 
\be
\Res_{X=0} \frac{1}{X}[2 \eta_2(X, q, z) -  \tilde{\eta} (X, q, z)] =1. 
\ee

Surprisingly, without concerning about Calabi-Yau conditions and mirror symmetry, the formalism in this section can be actually applied to a much more general curve 
\be \label{generalcurve} 
e^{\hat{p}} +z e^{-m\hat{x}-\hat{p}} =r(X), 
\ee
where the function $r(X)$ can be a general rational function of $X=e^{\hat{x}}$. We can also relax the constrain $m>0$ and allow it to be any integer. The mirror curve (\ref{mncurve}) is a special case of $r(X)=1-X$. We can still similarly solve for the function $V(X)$ perturbatively. For the case of $r(X)=1-X$, the leading term $\frac{1}{ r(X)} $ does not contribute in the residue calculations. However this is not true for a general rational function $r(X)$. It is more convenient to directly normalize $V(X)$ and also define a function $s(X)$ as 
\be
\tilde{V}(X)\equiv r(q^{-\frac{1}{2}}X) V(q^{-\frac{1}{2}}X), ~~~ s(X)\equiv  1/[X^{m} r(q^{\frac{1}{2}}X)r(q^{-\frac{1}{2}}X)].   
\ee
Here we also shift $X$ by a factor of $q^{-\frac{1}{2}}$ in the definition of $\tilde{V}(X)$, so that the difference equation looks much simplified
\be \label{tildeV} 
\frac{1}{\tilde{V}(X)}  +\tilde{V}(qX) s(X) z  =1. 
\ee
We can recursively compute the small $z$ expansion of $\tilde{V}(X)$ and its logarithm. The explicit expression  up to a few orders is 
\be
\ba
 \label{generalV} 
\log(\tilde{V}(X)) &=s(X) z   + \frac{1}{2} s( X) [s( X) + 2s(q X)]  z^2 + \frac{1}{3} s(X) [ s(X)^2 + 3 s(X) s(q X) \\ & + 3 s(q X)^2+ 3s(qX) s(q^2 X)] z^3 +\mathcal{O}(z^4).  
\ea
\ee
The residue is then defined by 
\be
 \Pi_{s(X)} =  \Res_{X=0} \frac{\log[  \tilde{V}(X)] }{X} =  \Res_{X=0} \frac{\log[ r(X) V(X)] }{X} . 
\ee
Here a shift $X$ by a factor does not affect the constant term, so the residue is the same in the above equation. 

On the other hand, the derivation of the TBA-like equation for a function $\eta_{s(X)}$ also goes through smoothly for this class of curves, we have 
\be 
\ba  \label{firstTBAgeneral}
1+ s(X) [\eta_{s(X)} (qX) +\eta_{s(X)}(X)] [ \eta_{s(X)}(q^{-1} X) +\eta_{s(X)}(X)] z =\eta_{s(X)}(X)^2 .
\ea
\ee
We can also explicitly compute the small $z$ expansion up to a few orders 
\be
\ba
\eta_{s(X)} (X) &= 1 + 2s(X) z+ 2 s(X) [s(X)+ s(q^{-1}X) + s(qX)  ]z^2 + 2 z^3 s(X) [ s(X)^2  \\ & +  2 s(X) (s(q^{-1}X)+ s(q X) )  + s(q^{-2}X) s(q^{-1} X) + 
s(q^{-1}  X) ^2   \\ & + s(q^{-1} X) s(q X)  + 
   s(q X) ^2 + s(q X) s(q^2 X)  ] + \mathcal{O}(z^4). 
\ea   
\ee 

We can now check that the following relation is still valid as a formal $z$ power series for various functions $s(X)$  
\be  \label{generalcompare} 
\Res_{X=0} \frac{1}{X} \eta_{s(X)} (X, q, z) = 1+2  \theta_z \Pi_{s(X)} .
\ee
This is true for any function $s(X)$  as long as it has a Laurent expansion around $X\sim 0$. The dependence of the function $s(X)$ on $q$ does not affect the relation, since the coefficients of its Laurent expansion are regarded as free parameters.

It is not difficult to explicitly check the relation  (\ref{generalcompare})  up to a finite order in $z$ expansion. We note for any function $f(X)$ with a Laurent expansion around $X\sim 0$, the constant term remains the same with a scaling of $X$ by any constant $a$, i.e. we have  $\Res_{X=0} \frac{1}{X} [f(aX) -f(X)]= 0$.   We can then for example show that the relation (\ref{generalcompare})  holds up to second order using e.g. $\Res_{X=0} \frac{1}{X} s(X)[ s(qX)- s(q^{-1} X)]= 0$.

Suppose around $X\sim 0$, we have the leading order $s(X)\sim X^d$. The residues in (\ref{generalcompare}) are generically non-trivial for $d<0$. On the other hand, if $d>0$, it is easy to see that $X=0$ is a zero point of both $ \tilde{V}(X)$ and $\eta_{s(X)} (X)$  at order $z$. Further, it is not difficult to show recursively that $X=0$ remains a zero point  of both functions in the higher order $z$ perturbations, so both sides of the equation (\ref{generalV}) are  trivially 1 in this case. 

The situation is not completely trivial but still much simplified for the critical case of $d=0$. It is not difficult to show recursively that the functions  $\eta_{s(X)}(X)$ and $\tilde{V}(X)$ have no pole around $X\sim 0$ and the constant terms do not vanish but are independent of $q$, i.e. purely classical. Suppose $s(X) =c +\mathcal{O}(X)$ where $c$ is a non-zero finite constant. It is easy to check the classical limit with $X=0$ and explicitly solve the functions with the appropriate choice of the square root branches. We find 
\be
 \eta_{s(X)}(X) |_{X=0} =\frac{1}{\sqrt{1-4cz} },  ~~~ \tilde{V}(X)|_{X=0} = \frac{1-\sqrt{1-4cz}}{2cz}. 
\ee
After a simple algebraic calculation with $z$ derivative, we see both sides of the equations (\ref{generalcompare}) are exactly $\frac{1}{\sqrt{1-4cz} }$, i.e. the same.

\section{The $\mathcal{O}_{m,1}$ operators: second method}   \label{sec4}

In this section, we provide another method to derive the TBA-like equation for the quantum periods of the curves (\ref{mncurve}) of the $\mathcal{O}_{m,1}$ operator in the previous section \ref{sec3}. This approach use the formulas in \cite{Kashaev:2015} for writing the $\mathcal{O}_{m,n}^{-1}$ operator in terms of Faddeev's quantum dilogarithm, in similar fashion as in (\ref{spectral}, \ref{spectraloperator}). The notable difference is that the momentum operator is shifted by a constant. The derivations of the TBA-like equations with this shift have been worked out in \cite{Okuyama:2015pzt}, following the earlier original paper \cite{Tracy:1995ax}. The paper \cite{Okuyama:2015pzt} also already provided some tests of the proposal of \cite{Hatsuda:2013oxa} for the $\mathbb{P}^2$ model  in a semiclassical small $\hbar$ expansion. Our focus will be a simpler small $z$ expansion which keeps $\hbar$ finite similar as in the previous section. Our method will be easily generalized to the cases of  $\mathcal{O}_{m,1}$ operators. 

We provide a brief review of some crucial technical steps in \cite{Okuyama:2015pzt}, and extract the relevant TBA-like difference equations for our purpose.  For simplicity, we give most details for the $\mathbb{P}^2$ model, i.e. $m=1$ case, and then just present the results for general cases which are similarly derived. As we will see, although the TBA-like equations look very different from those in the previous section, the relevant residues are related to the same quantum period.  

After a change of coordinates $x,p$, as well as a similarity transformation of the inverse of the operator (\ref{OMN}), which were discussed in details in \cite{Kashaev:2015, Okuyama:2015pzt}, the integral kernel becomes the following expression 
\be
\bra x_1 | \rho_{m,n} | x_2\ket = \frac{ u_{m,n}(x_1)^{-\frac{1}{2}} u_{m,n}(x_2)^{-\frac{1}{2}} }{ 2(m+n+1)\hbar \cosh(\frac{\pi(x_1-x_2)}{(m+n+1)\hbar}+\frac{i (m-n+1)\pi}{2(m+n+1)} )}, 
\ee
where the potential $u_{m,n}(x)$ can be described by Faddeev's quantum dilogarithm 
\be
u_{m,n}(x)=e^{-\frac{m}{m+n+1}x} |\Phi_{b}(\frac{x-i\hbar(m+1)/2}{2\pi b})|^2 ,
\ee
with the parameter $b:=\sqrt{\frac{(m+n+1)\hbar}{2\pi}}$. Though the original operator (\ref{OMN}) is symmetric in $(m,n)$, it is obscured by the quantum dilogarithm and furthermore not respected by the similarity transformation. Using the quasi-periodic relation of the quantum dilogarithm in Appendix \ref{Qdilogorithm}, it is straightforward to obtain the following product formula 
\be
\ba \label{productformula}
\prod_{k=-(m+n)/2}^{(m+n)/2}u_{m,n}(q^{k}X) 
&=X^{-m}\frac{\prod_{k=-m-\frac{n+1}{2}}^{-\frac{n+1}{2}}\Phi_{b}(\frac{x+i\hbar k}{2\pi b})}{\prod_{k=\frac{n+1}{2}}^{m+\frac{n+1}{2}}\Phi_{b}(\frac{x+i\hbar k}{2\pi b})} \\ &=X^{-m}\prod_{k=-m/2}^{m/2}(1+q^kX),
\ea
\ee
where the notations are $q=e^{i\hbar}, X=e^x$. We see that the quantum dilogarithms cancel out in the product, so the result is actually just an elementary function. This will be quite useful since we would like to have nice TBA-like equations involving only elementary functions.

Now we specialize to the $m=n=1$ case which is the local $\mathbb{P}^2$ model. The integral kernel can be written as 
\begin{eqnarray} \label{rhofunction}
	\rho (x_1,x_2) =\frac{\sqrt{E(x_1)E(x_2)}}{\alpha M(x_1)+\alpha^{-1}M(x_2)}, 
\end{eqnarray}
where the notations are 
\begin{align} \label{notation}
	\alpha&=e^{i \pi/6}, ~~~~~~~ \omega=-\alpha^{-2} =e^{2 i \pi/3}, \\
	E(x)&=\frac{1}{3\hbar}\frac{M(x)}{u(x)}, ~~~ M(x):=e^{\frac{2\pi x}{\hbar}}, ~~~ u(x):=u_{1,1}(x) .
\end{align}

We need to define some more notations. The resolvent operator is defined by the integral kernel $R(\kappa):=\frac{\rho}{1-\kappa \rho}$, and we split it into 3 parts
\begin{align}
	R_k(\kappa)&=\frac{\kappa^k \rho^{k+1}}{1-(\kappa \rho)^3},~~~~(k=0,1,2)\\
	R(\kappa)&=\sum_{k=0}^{2}R_k(\kappa).
\end{align}
The variable $\kappa$ of the generating function will be later properly identified with the complex structure parameter.  We also need to define a series of functions by recursion from an initial function 
\begin{align}
\phi_j (x):=\int_{-\infty}^{\infty} dx^{'} \frac{1}{\sqrt{E(x)}}\rho(x,x^{'})\sqrt{E(x^{'})}\phi_{j-1}(x^{'}), ~~~~~\phi_0(x)=1. 
\end{align}
Like $R_k(\kappa)$, we can also split the generating functions of $\phi_l$'s into 3 parts as 
\begin{align}
	\Psi_k:=\sum_{j=0}^{\infty} \kappa^{k+3j}\phi_{k+3j},~~~~~\bar{\Psi}_k:=\sum_{j=0}^{\infty} \kappa^{k+3j}\bar{\phi}_{k+3j},~~~~~k=0,1,2,
\end{align}
where $\bar{\phi}_k$ denotes the complex conjugate of $ \phi_k$.

From the definitions of the functions, it is straightforward to find the relations between $\phi_j$ to $ \rho^j $ \cite{Okuyama:2015pzt}. So the generating function $\Psi_k$ is related to $R_k$ as 
\begin{eqnarray}\label{Rkresolvent}
	R_k(x,x^{'})=\alpha^{-1}\frac{\sqrt{E(x)E(x^{'})}}{M(x)-\omega^{k+1} M(x^{'})}\sum_{r=0}^{2}\omega^r\Psi_r(x)\bar{\Psi}_{k-r}(x^{'}).
\end{eqnarray}

In our consideration the integral kernel $\rho$ is a trace class operator, so the function of $R_k$ is well defined, in particular with no singularity at the diagonal elements $x=x^\prime$. For the case $k=2$, the denominator in (\ref{Rkresolvent}) is zero when $x=x^\prime$, so we have a constrain for the $\Psi_r$ functions 
\begin{eqnarray}\label{Constrain}
	\sum_{r=0}^{2}\omega^r\Psi_r(x)\bar{\Psi}_{2-r}(x)=0.
\end{eqnarray} 
In the following discussions we will only need to use the diagonal function well defined by $R_k(x) := \lim_{x^{\prime}\rightarrow x} R_k(x,x^{'})$. 

Now we will build up the TBA difference equations, from the Lemma 2 in \cite{Tracy:1995ax}, we have the difference equation about $\Psi_k$
\be
\ba
\label{Phidifference}
	& \Psi_k(x+i\hbar)-\Psi_k(x-2i\hbar)=\frac{\kappa\omega^{-1/2}}{u(x)}\Psi_{k-1}, \\
	& \bar{\Psi}_k(x-i\hbar)-\bar{\Psi}_k(x+2i\hbar)=\frac{\kappa\omega^{1/2}}{u(x)}\bar{\Psi}_{k-1},
\ea
\ee
where $k$ is modulo 3. The above relations are not enough to build TBA-like difference equations. We need some other important relations called ``quantum Wronskian relations", which are
\be
\ba
\label{Wronskian}
&\sum_{r=0}^{2}\omega^r\Psi_r(x+\frac{1}{2}i\hbar)\bar{\Psi}_{3-r}(x-\frac{1}{2}i\hbar)=1, \\&	
\sum_{r=0}^{2}\omega^r\Psi_r(x-i\hbar)\bar{\Psi}_{3-r}(x+i\hbar)=1. 
\ea
\ee
These relations can be rigorously proven along the lines of \cite{Tracy:1995ax}, and assuming the right analytic properties of the relevant functions, a simple heuristic derivation was also given in \cite{Okuyama:2015pzt}. 

We can now define some $\eta$ functions by inserting some phase factors into the quantum Wronskian relations
\begin{align}
\eta_k&:=\omega^{3/2}\sum_{r=0}^{2}\omega^{r-\frac{k+1}{2}}\Psi_r(x-i\hbar)\bar{\Psi}_{k+1-r}(x+i\hbar), \nonumber\\
\tilde{\eta}_k&:=\omega^{3/2}\sum_{r=0}^{2}\omega^{r-\frac{k+1}{2}}\Psi_r(x+\frac{1}{2}i\hbar)\bar{\Psi}_{k+1-r}(x-\frac{1}{2}i\hbar).
\end{align}
where $\eta_2=\tilde{\eta}_2=1$ according to the quantum Wronskian relations, so here we only consider the cases of $k=0,1$.

Using the relations (\ref{Phidifference}) and (\ref{Rkresolvent}), as well as some heuristic arguments similar to the derivation of the quantum Wronskian relations in \cite{Okuyama:2015pzt}, one can derive the follow difference equations about $\eta_k$
\begin{align}
	\frac{\sin(3/2\hbar\partial_x)}{\sin(1/2\hbar\partial_x)}\eta_k = 6\hbar \kappa \sin(\frac{k+1}{3}\pi)R_k,
\end{align}
where the operator is simply a sum of difference functions $\frac{\sin(3/2\hbar\partial_x)}{\sin(1/2\hbar\partial_x)}f(x) = f(x-i\hbar)+f(x) + f(x+i\hbar)$. 

Combining the $k=0,1$ equations properly, we get the first TBA-like difference equation
\begin{eqnarray}\label{TBA1}
	-\kappa u(x)^{-1}\mathcal{R}(x)=\eta(x-i\hbar)+\eta(x)+\eta(x+i\hbar),
\end{eqnarray}  
where we introduce some more new functions 
\be
\ba
\eta&:=-\sum_{k=0}^{1}\omega^{3/2k}\eta_k = \eta_1-\eta_0, \\
\mathcal{R}& :=\Psi\bar{\Psi}=3\sqrt{3}\hbar u(x)(R_0-R_1) \\
\Psi&:=\sum_{k=0}^{2}\omega^{-k}\Psi_k,~~~~~\bar{\Psi}:=\sum_{k=0}^{2}\omega^k\bar{\Psi}_k.	
\ea
\ee
The residue of the function $\eta(x)$ will be compared with quantum periods. Here we have chosen a proper normalization and sign which would be convenient for the comparison. 

Using the quantum Wronskian relations, we can get the next TBA-like difference equation
\be
\ba  \label{TBA2}
	&\mathcal{R}(x+i\hbar)\mathcal{R}(x)\mathcal{R}(x-i\hbar)   \\
	&= (1 - \eta(x))(1+\eta(x)+\eta(x-i\hbar))(1+\eta(x)+\eta(x+i\hbar)).
\ea
\ee

There is another TBA-like difference equation corresponding to $R_2$, and we do not list it here since we need not it for our purpose. Substituting the equation (\ref{TBA1}) into (\ref{TBA2}) and setting $ z:=\kappa^{-3} $, we can get the TBA-like difference equation 
\be
\ba 
\label{TBAp2}
  & (1-\eta(X))(1+\eta(X)+\eta(q^{-1}X))(1+\eta(X)+\eta(q X)) \\
  & = -z u(q^{-1}X) u(X) u(q X) A(q^{-1}X) A(X) A(qX) ,  \\
  & {\rm with} ~~~A(X):=\eta(q^{-1}X)+\eta(X)+\eta(q X).
\ea
\ee
The above TBA-like equation is valid for any function $u(x)$ as long as $\rho$ is of trace class. In our particular example, we can use the product formula (\ref{productformula}) for $m=n=1$, which is 
\be
u(q^{-1}X) u(X) u(q X)=X^{-1}(1+X q^{1/2})(1+Xq^{-1/2}). 
\ee
So we finally obtain a TBA-like difference equation for the local $\mathbb{P}^2$ model with only elementary functions 
\be
\ba 
\label{TBAp2a}
  & (1- \eta(X))(1+\eta(X)+\eta(q^{-1}X))(1+\eta(X)+\eta(q X)) \\
  & = -z X^{-1}(1+X q^{1/2})(1+Xq^{-1/2}) A(q^{-1}X) A(X) A(qX) ,  
\ea
\ee
We can solve the difference equation (\ref{TBAp2}) as the perturbative series of $z$. We are interested in the solution with 1 as the leading term, which is 
\begin{align}
	\eta(X,q,z)=1+3(q^{-1/2}+q^{1/2}+X+X^{-1})+\mathcal{O}(z^2). 
\end{align}
We can now compare the residue with the relevant quantum period in (\ref{Pim}). We check the relation perturbatively 
\begin{align}
\Res_{X=0} \frac{1}{X} \eta(X, q, z) = 1+3 \theta_z \Pi_1,
\end{align}

Next we consider the operator $\mathcal{O}_{m,n}$ in (\ref{OMN}) for general $m,n$. Follow the similar procedure as in \cite{Okuyama:2015pzt}, we can derive the relevant TBA-like equations. However, only in the case of $n=1$, we can readily solve the TBA-like equation in perturbative series. So for our purpose for now we focus on the operators $\mathcal{O}_{m,1}$, which is also the situation for the first method in the previous section \ref{sec3}. In this case, the resolvent kernel is split into $m+2$ parts. After some calculations, we find the following TBA-like difference equation for a properly defined function $\eta_m(X)$
\be
\ba 
&(1-\eta_m(X))\prod_{i=-m}^{0}B(q^{i}X) \
=-z \prod_{k=-(m+1)/2}^{(m+1)/2}u(q^{k}X) \prod_{i=-(m+1)}^{0}A(q^{i}X) , \\
	&{\rm with}~~~ A(X):=\sum_{i=0}^{m+1}\eta_m(q^{i}X), ~~~
	B(X):=1+\sum_{i=0}^{m}\eta_m(q^iX).
\ea
\ee
Again using the product formula (\ref{productformula}), we arrive at a nice TBA-like equation with only elementary functions 
\be
\ba \label{TBAm}
&(1-\eta_m(X))\prod_{i=-m}^{0}B(q^{i}X) \
=-z X^{-m}\prod_{k=-m/2}^{m/2}(1+q^kX) \prod_{i=-(m+1)}^{0}A(q^{i}X) . 
\ea
\ee
For the special case $m=1$, this equation for $\eta_1(X):=\eta(X)$ reduces to (\ref{TBAp2a}) for local $\mathbb{P}^2$ model. We should note that the $\eta_m(X)$ function here is completely different from those in (\ref{firstTBA}) in the previous section \ref{sec3}. Without confusing the notations, we still use the same symbol, since their residue are both related to the same quantum period and they are self evidently defined by their own TBA-like equations.

We solve the equations for some small $m$ numbers perturbatively, and list some results here 
\be
\ba
	\eta_2(X,q,z)&=1+4(1+q^{-1}X)(1+X)(1+qX)X^{-1}z+\mathcal{O}(z^2), \\
	\eta_3(X,q,z)&=1+5(1+q^{-\frac{3}{2}}X)(1+q^{-\frac{1}{2}}X)(1+q^{\frac{1}{2}}X)(1+q^{\frac{3}{2
	}}X)X^{-1}z+\mathcal{O}(z^2), \\
	\eta_4(X,q,z)&=1+6(1+q^{-2}X)(1+q^{-1}X)(1+X)(1+qX)(1+q^2X)X^{-1}z+\mathcal{O}(z^2). 
\ea\ee
We take residue for $X$ at 0, and check perturbatively the the following relations with quantum periods in (\ref{Pim}) for some small $m$ numbers 
\begin{eqnarray} \label{compare2}
	\Res_{X=0}\frac{1}{X}\eta_m(X,q,z)=1+(m+2) \theta_z\Pi_{m}. 
	\end{eqnarray}

\section{Discussions} \label{sec5}

In the WKB expansion of small $\hbar$ parameter, the quantum corrections to classical periods can be expressed exactly as differential operators acting on the classical periods. As exemplified in our paper \cite{Huang:2020yli}, these differential operators may be derived more easily using the corresponding TBA-like equations. In the class of geometries of $\mathcal{O}_{m,1}$ operators, the TBA-like equations (\ref{firstTBA}) in Section \ref{sec3} are similar to those of $\mathbb{P}^1\times \mathbb{P}^1$ geometry in \cite{Huang:2020yli}. For example, in the classical limit $\hbar\rightarrow 0$, the TBA-like equation becomes a simple quadratic equation with no linear term, which is simpler than the analogous equation in the calculations of the classical period.  One further simplification of TBA-like equation (3.12) is that it is symmetric with $\hbar\rightarrow -\hbar$, so the WKB expansion has only even powers of $\hbar$. On the other hand, the WKB expansion of quantum wave function $\log(V(X))$ defined in (\ref{quadratic3.3}) does have odd powers of $\hbar$, which turn out to be total derivatives and only vanish after taking residue. We expect that it should be a straightforward exercise to work out the differential operators of quantum corrections with no surprise. If interesting circumstances and applications arise in the future, one may use the TBA-like equations (\ref{firstTBA}) in Section \ref{sec3} to more easily work out the details of the differential operators for quantum corrections.

However, the TBA-like equations (\ref{TBAm}) in Section \ref{sec4} are more complicated. In the classical limit, it becomes a degree $m+2$ polynomial equation for $\eta_m$. For $m=1,2$ one can still have analytic solution, as also considered in \cite{Okuyama:2015pzt}. However it does not generically have an algebraic solution for $m>2$. So it seems that the TBA-like equations (\ref{TBAm}) are only suitable for calculations in small $z$ perturbation, but not much in small $\hbar$ perturbation.

In this paper we focus on a simple A-cycle where the period integral reduces to a residue. For the other more complicated B-cycles, it is usually difficult to directly perform the integrals exactly in $\hbar$.  A better approach is to do a WKB expansion of small $\hbar$ parameter, then the quantum corrections are determined by differential operators which are the same for all cycles. In this way, one can also compute the quantum B-period in terms of differential operators acting on classical B-period, though only perturbatively in $\hbar$ expansion.

It is well known that many 5d supersymmetric gauge theories are geometrically engineered by toric Calabi-Yau geometries, e.g. the 5d pure $SU(2)$ gauge theory is engineered by local $\mathbb{P}^1\times \mathbb{P}^1$ geometry. In the 4d limit, the quantum operator reduces to that of a non-relativistic particle with the standard quadratic kinetic term $\hat{p}^2$.  The connections between quantum periods of the 4d Seiberg-Witten-like curves  and certain TBA-like equations have appeared many times in the literature, see e.g. \cite{Kozlowski:2010tv, Meneghelli:2013tia, Grassi:2018bci, Ito:2018eon}. In the 4d cases, the relation between quantum periods and TBA-like equations appear to be mostly well established by now, though it seems the TBA-like equations are usually a reformulation which does not necessarily simplify the calculations much. In our case, the appearance of the factor of $\cosh(\hat{p})$ in the quantum operator e.g. (\ref{spectraloperator}) is crucial, and we are also taking residue with the exponentiated parameter $X=e^x$. So it seems the relation studied here is intrinsically 5 dimensional, and we are not aware a simple 4d limit. It would be interesting to study this issue further.

We should mention that in both 4d and 5d supersymmetric gauge theories, instead of using the Seiberg-Witten curves, there is yet another method to determine the quantum mirror maps by a calculation of the vacuum expectation value of 4d chiral operators or 5d Wilson loop operators in the NS limit of the $\Omega$ background, explained in some details in e.g.  \cite{Grassi:2018bci}. It would be interesting  to study the potential relation of the present work to this different method.

The most intriguing feature of this paper is that the two different approaches in Sections \ref{sec3} and \ref{sec4} give rise to entirely different TBA-like equations and perturbative solutions for the class of Calabi-Yau geometries. However, in both cases, the residues of the perturbative solutions of TBA-like equations are related to the same quantum period as in (\ref{compare1}, \ref{compare2}). In one particular case, namely the geometry of $\mathcal{O}_{2,1}$ operator, we even have three different TBA-like equations due to a geometric equivalence of the local $\mathbb{F}_0$ and  $\mathbb{F}_2$ Calabi-Yau models, see e.g. (\ref{F2equation}) for the extra equation. In this sense, our paper provides multiple different realizations of the same geometry. It would be interesting to study how the residues in these different looking TBA-like equations may be directly related without using quantum periods as a connecting hub.  

While the first method in Section \ref{sec3} applies more generally to a larger class of curves (\ref{generalcurve}), the second method in Section \ref{sec4} appears to work more specifically for the $\mathcal{O}_{m,1}$ operators.  Certainly, it would be interesting to further generalize the results to more Calabi-Yau geometries and consider a bigger moduli space instead of the one-parameter space in this paper.  In the case of local Calabi-Yau geometries with multiple A-periods, it is found that their quantum corrections are described by the same differential operators \cite{Huang:2020neq}. It seems that the general formalism here would need to be much improved to find all the TBA-like equations for the quantum periods of the different A-cycles of a Calabi-Yau geometry.

It may be interesting to follow the steps in \cite{Kallen:2013qla} to provide a more rigorous derivation of the relation between quantum periods and TBA-like equations discussed here. However,  for the first method in Section \ref{sec3}, as we mentioned, the spectral theory is merely used as a formal trick, and may not be well behaved since the integral kernel may not be of trace class. So it seems unlikely at least in this case that one may establish a rigorous link using the spectral theory, and some new and more unifying approaches may be needed.

\vspace{0.2in} {\leftline {\bf Acknowledgments}}

We thank Sheldon Katz, Albrecht Klemm, Yuji Sugimoto, Xin Wang, Di Yang, Pei-xuan Zeng for helpful discussions, and/or stimulating collaborations on related papers. This work was supported in parts by the national Natural Science Foundation of China (Grants No.11675167 and No.11947301).

\appendix

\section{Quantum dilogarithm} \label{Qdilogorithm}
We provide some useful properties of Faddeev's quantum dilogarithm, defined e.g. in \cite{Faddeev:1993rs}.  First, the (conventional) quantum dilogarithm is defined by 
\be
\phi(x) \equiv (x; q)_{\infty} = \prod_{n=0}^{\infty} (1-x q^n), ~~~~ |q|<1
\ee 
The Faddeev's quantum dilogarithm can be defined by 
\be 
\Phi_b (x) = \frac{ (e^{2\pi b (x+c_b)} ; e^{2\pi i b^2})_{\infty} }{ (e^{2\pi b^{-1} (x-c_b)} ; e^{-2\pi i b^{-2}} )_{\infty} } , 
\ee
where $c_b=\frac{i}{2}(b+b^{-1})$. The formula is well defined for $\Im(b^2)>0$ but can be analytically extended to all values of $b$ except for $b^2$ a non-positive real number. In this paper we simply call this function $\Phi_b (x) $ the quantum dilogarithm.  

A useful property is the unitarity relation 
\be
\Phi_b(x)^* \Phi_b(\bar{x}) =1. 
\ee
In particular, for a real number $x$ we have $|\Phi_b(x)|=1$, which is just a pure phase. 

The following quasi-periodic relations are also frequently used in this paper
\be
\ba
& \frac{\Phi_b(x-ib)}{\Phi_b(x) } = 1-e^{2\pi b(x- c_b)},  \\ 
& \frac{\Phi_b(x-ib^{-1})}{\Phi_b(x) } = 1-e^{2\pi b^{-1}(x-c_b)} . 
\ea
\ee

\addcontentsline{toc}{section}{References}


\providecommand{\href}[2]{#2}\begingroup\raggedright\endgroup

\end{document}